# Depth dependent magnetization profiles of hybrid exchange springs

T. N. Anh Nguyen[1,2,a], R. Knut[3], V. Fallahi[4], S. Chung[1,5], S. M. Mohseni[1,6,7], Q. Tuan Le[1], O. Karis[3], S. Peredkov[8], R. K. Dumas[5], Casey W. Miller[9], and J. Åkerman[1,5,6]

[1]*Materials Physics, School of ICT, Royal Institute of Technology, Electrum 229, 164 40 Kista, Sweden*

[2]*Spintronics Research Group, Laboratory for Nanotechnology (LNT), Vietnam National University – Ho Chi Minh City (VNU-HCM), Ho Chi Minh City, Vietnam*

[3]*Department of Physics and Astronomy, Uppsala University, Box 516, 75120, Sweden*

[4]*Department of Optics and Laser Engineering, University of Bonab, 5551761167, Bonab, Iran*

[5]*Department of Physics, University of Gothenburg, 412 96 Gothenburg, Sweden*

[6]*NanOsc AB, Electrum 205, 164 40 Kista, Sweden*

[7]*Department of Physics, Shahid Beheshti University, Tehran, Iran*

[8]*Max-lab, Lund University, Box 118, 22100 Lund, Sweden*

[9]*Department of Physics, University of South Florida, 4202 East Fowler Avenue, Tampa, FL 33620, USA*

We report on the magnetization depth profile of a hybrid exchange spring system in which a Co/Pd multilayer with perpendicular anisotropy is coupled to a CoFeB thin film with in-plane anisotropy. The competition between these two orthogonal anisotropies promotes a strong depth dependence of the magnetization orientation. The angle of the magnetization vector is sensitive both to the strength of the individual anisotropies and to the local exchange constant, and is thus tunable by changing the thickness of the CoFeB layer and by substituting Ni for Pd in one layer of the Co/Pd stack. The resulting magnetic depth profiles are directly probed by element specific x-ray magnetic circular dichroism (XMCD) of the Co, Fe, and Ni layers located at different average depths. The experimental results are corroborated by micromagnetic simulations.

PACS numbers: 75.30.Gw, 75.30.Et, 75.70.-i, 78.20.Ls, 75.70.Cn

---

[a]Author to whom correspondence should be addressed. Electronic mail: anhntn@kth.se.

The phenomenon of spin transfer torque (STT) [1-3], in which a spin-polarized current transfers angular momentum to a magnetic layer, has brought about novel applications such as spin torque oscillators (STOs) [4-7] and spin transfer torque magnetoresistive random access memory (STT-MRAM) [8-10]. While initially based on magnetic materials with in-plane (IP) magnetic anisotropy, the realization that such materials lead to unnecessarily high STT-MRAM switching currents, poor memory retention, poor scalability [11], and high-field operation of STOs [12], there is now a rapidly growing interest in fabricating STT devices based on perpendicular magnetic anisotropy (PMA) materials. Recent tailoring of PMA materials and their interfaces have demonstrated low switching currents, high switching speed, good thermal stability, future scalability [9,13-15], and low- to zero-field operation of STOs [16-20].

Building upon these successes, the natural extension of using PMA materials is to also investigate the potential of devices in which the magnetization is *tilted* with respect to the surface normal. Such materials allow for additional control of the magnetization dynamics in magnetic nanostructures [17,21-23], and hint at yet improved STT-MRAM switching behavior and thermal stability [24-29]. For STOs, tilted materials offer a route to improve their microwave generation properties, both in terms of higher output power and low- to zero-field operation [17,21-23,28,30-32].Recently, tilted materials have also been shown to have potential for current-driven domain wall motion [33]. The influence of a tilted anisotropy is stronger than simply tilting the applied field [34] as a mere 5 degree misalignment between the free and the fixed layer in magnetic tunnel junctions (MTJs) can reduce the switching current by 36%, the switching time by 30%, and improve the switching current distribution [35].

Materials with tilted anisotropies have been realized using collimated oblique sputtering [36], depositing multilayers on nanospheres [25], and exploiting crystallographic texture to control the magnetic easy axis in alloys such as (112)-textured $D0_{22}$MnGa (with a tilt angle of 36°), and (111) or (101)-$L1_0$FePt (with angles of 36° and 45°, respectively) [37-39]. Recently, an alternative and much more versatile approach was reported where exchange springs combining materials with out-of-plane (OOP) and IP anisotropies provide a wide and tunable range of tilt angles [40-42]. Using different thicknesses of the OOP and IP layers and different OOP-IP coupling strengths, the average tilt angle, the details of the magnetization profile, and even the damping, can be varied with great freedom. However, knowledge about the actual highly non-linear thickness dependent magnetization profile was only inferred indirectly using a combination of magnetometry and 1-dimensional micromagnetic modeling.

In this letter, we present a depth resolved x-ray magnetic circular dichroism (XMCD) study of the spin orientation in OOP/IP exchange springs where a digital Ni layer, inserted at various depths, is used as an additional local probe of the magnetization direction throughout the exchange spring. By taking advantage of the inherent elemental specificity of XMCD we are able to directly probe the magnetization orientation of the different elements in the OOP and IP layers, as well as that of the buried Ni layer, and provide the missing experimental piece of information of the magnetization profile in tilted exchange springs.

The tilted exchange springs have a top $Co_{40}Fe_{40}B_{20}$ (CFB) layer which is grown on a [Co/Pd] multilayer. CFB was chosen for being magnetically soft and the material of choice when exchange springs are implemented in MTJs with MgO barriers. All film stacks were deposited at room temperature on thermally oxidized Si substrates using a confocal magnetron sputtering system in a chamber with a base pressure below $3 \times 10^{-8}$ Torr. The films were grown on Ta (10 nm)/Pd (3 nm) seed layers to improve the PMA of the Co/Pd [43,44]. The CFB was deposited as a wedge with thicknesses of 0.5 to 1.75 nm by oblique deposition from a stoichiometric target. Finally, a 2 nm-thick Ta capping layer protects the CFB against oxidization. To tune the properties of the exchange spring we prepared two main sample series. Series A has a fixed CFB thickness (1.75 nm) with a Co/Ni bilayer at different positions (n) within the PMA stack: [Co(0.5 nm)/Pd(1.8 nm)]$_{4-n}$/Co(0.5 nm)/Ni(1 nm)/[Co (0.5 nm)/Pd(1.8 nm)]$_n$/CFB(1.75 nm). In series B only the CFB thickness ($t_{CFB}$) was varied: [Co (0.5 nm)/Pd(1.8 nm)]$_2$/Co(0.5 nm)/Ni(1 nm)/[Co (0.5 nm)/Pd(1.8 nm)]$_2$/CFB($t_{CFB}$). To verify that the insertion of a Ni layer, and its exact position, did not vary the overall properties of the PMA layer, an additional series of control samples were deposited: [Co/Pd]$_5$-CFB(0.5 nm) and [Co/Pd]$_{4-n}$-Co/Ni-[Co/Pd]$_n$-CFB(0.5 nm) (n= 0, 1, and 2). OOP and IP hysteresis loops were measured using an alternating gradient magnetometer (AGM) at room temperature.

To verify that the Ni insertion layer has a negligible influence on the PMA of the Co/Pd stack, OOP hysteresis loops were measured of the control series (Fig.1). Substituting Ni for Pd results in a reduction of $H_C$, expected for exchange coupled composite structures [45,46]. When the Ni insertion layer moves deeper into the Co/Pd MLs, (Co/Pd)$_5$ is broken into two parts with the soft Co/Ni in between, resulting in a monotonic decrease of $H_C$ from the control sample's value of ~680 Oe down to ~480 Oe. Even with this reduction in $H_C$, a strong PMA with a well-defined square loop and narrow switching field distribution is always observed. The IP loops (not shown here) reveal an IP saturation field of about 1 T for all samples, which further confirms a strong PMA in these exchange spring MLs.

Fig. 2(a) shows the schematic illustration of a wedge-type [Co/Pd]-[Co/Ni]-[Co/Pd]-CFB ML stack. The IP and OOP hysteresis loops for samples with $t_{CFB}$ = 0.5, 0.85, 1, 1.25 and 1.75 nm, are shown in Figs. 2(b) and 2(c), respectively. The data clearly reveal that the competition between the in-plane magnetic anisotropy (IMA) of the CFB layer and the PMA of the [Co/Pd]$_2$-[Co/Ni]-[Co/Pd]$_2$ stack has a dramatic effect on the magnetization reversal as $t_{CFB}$ is increased. A significant OOP remanence is retained for $t_{CFB}$=0.5 nm, consistent with rigid coupling of the CFB layer to the PMA stack during reversal. However, as $t_{CFB}$ is increased, the IMA contribution of the CFB layer begins to dominate, leading to a gradually reduced OOP remanence and an increased OOP saturation field. The complementary trends are also observed for the IP loops: a clear decrease in the IP saturation field is observed as $t_{CFB}$ is increased. In fact, for $t_{CFB}$=1.75 nm, the two competing anisotropies have become comparable in size and the IP and OOP loops show a comparable reversal behavior.

XMCD investigations were carried out at the synchrotron facility MAX-lab (beamlines I1011 and D1011) in Lund, Sweden, with 90% and 75% circularly polarized light, respectively. All samples were fully magnetized in an OOP field and then measured in remanence using total electron yield. The XMCD spectra were taken at the $L_3$ edges of Ni, Co, and Fe at varying angles between the incident x-ray and the remanent magnetization of the samples. The angle dependent asymmetries are then used to calculate the average magnetization direction of the respective elements. Fig.3 shows the Fe $L_3$ and Ni $L_3$ asymmetries as a function of the angle between the incidents circularly polarized x-rays and the surface normal. The asymmetry is proportional to the projection of the spin magnetic moment on the direction of the incident x-rays, and hence follows a sinusoidal form. The data have been fitted by the function cos(x-θ)*S(x) (dashed lines), where x is the angle between surface normal and the incident light and θ is the angle of the magnetization relative to the normal. S(x) corrects for saturation effects, as described by Nakajima et al. [47]. The direction of the average magnetic moment corresponds to the peak value of the asymmetry. Hence OOP magnetization corresponds to a maximum asymmetry at 0 degrees. However, it is easier to determine the zero crossing of the asymmetry, which corresponds to the angle orthogonal to the angle of average magnetization. Since the zero crossing for Fe asymmetry goes closer to 0 degrees with increasing n, the CFB layer is becoming more IP as the Ni layer goes deeper in the stack. However, the zero crossing for Ni moves away from 0 degrees with increasing n, and hence obtains a more OOP like character.

The XMCD asymmetry at the $L_3$ edge is not only sensitive to the spin magnetic moment but also to the magnetic dipole term and orbital magnetic moment [48,49]. We find that the orbital moment anisotropy will affect the derived magnetization angle with less than 1 degree for Ni and Fe and we have therefore ignored this effect [50]. The magnetic dipole term can, in some of our geometries, affect the derived magnetization angle of Fe and Ni up to 3 degrees and has therefore been accounted for [51]. We have also made corrections due to deviations between the remanent magnetization direction and the plane studied by the angular scans.

Fig. 4(a) illustrates the CFB thickness dependence (left) and Ni depth dependence (right) of the tilted angles in Fe, Co, and Ni. For very thin CFB samples ($t_{CFB}$<1 nm), the XMCD asymmetries show that the Co and CFB layers all maintain perpendicular anisotropy due to the rigid coupling between very thin CFB and PMA stack. However, for thicker CFB ($t_{CFB}$>1 nm), the stronger IP anisotropy of CFB causes the magnetic moments of the Fe and Co to tilt. The magnetic moment of Fe within the thickest CFB layer is tilted 73° due to the IP anisotropy becoming dominant. A deep Ni layer (n=2) remains OOP at all times but for Ni closer to the CFB a tilt is observed. Fig. 4(a) (right) clearly shows that the CFB and Ni are directly exchange coupled for n=0 (Ni depth=3.75 nm) and therefore exhibit the same angle of 43 degrees.

For limited cases, additional depth information can be obtained from the average angle of the Co magnetization. For example, the magnetization direction for almost all Co layers in sample CFB(1.75 nm)/n=2 can be directly estimated by assuming a strong exchange coupling between Ni and Co layers. Only the top Co/Pd layers is undetermined and can hence be estimated by measuring the Co asymmetry, as shown in Fig. 4(b). We have used the electron escape depths $\lambda_e^{Co} = \lambda_e^{Ni} = \lambda_e^{Pd}$ =2.5 nm and $\lambda_e^{Fe}$=1.7 nm [47,52] while for CFB, a combination of $\lambda_e^{Fe}$ and $\lambda_e^{Co}$ gives $\lambda_e^{CFB}$=1.86 nm. Both the magnetic dipole ($m_D$) and the orbital moment ($m_{orb}$) asymmetry are strong for thin Co layers and each need to be included to obtain proper angles of the Co spin magnetic moment. We used $m_D^\perp$ =0.43 for the Co layers in the PMA stack and $m_D^\perp$ =0.143 for the top CFB layer [53]. The orbital moment used for Co layers are $m_{orb}^\perp$ =0.25 and $m_{orb}^{//}$ =0.03 while $m_{orb}^\perp$=0.18 and $m_{orb}^{//}$ =0.1 for the top CFB layer [54]. The results are shown in Fig. 4(b) where the black solid line is the sum of all Co asymmetry contributions fitted to the experimentally obtained asymmetries (blue solid circles) and the green and purple lines correspond to the spin magnetic moment for the different Co layers. The 1st and 2nd Co layers correspond to the top two [Co/Pd] layers and the 3rd Co layer corresponds to the Co/Ni layer which is assumed to be strongly exchange coupled to the 2nd Co layer. The red solid line is the sum of all the spin contributions, which clearly illustrates the strong contribution from $m_D$ and $m_{orb}$ since it is distinct from the total asymmetry (black solid line). The spin moment direction of the 1st Co layer (green line) was a fitting parameter that gave the best fit for 60 degrees, which is in accord with the magnetization gradient calculated below.

Micromagnetic simulations that quantitatively determine the magnetization tilting within the various magnetic layers are fully consistent with the XMCD results. The calculations were based on a one-dimensional micromagnetic model. The magnetic configuration of each layer was calculated by minimizing the system's Gibbs free energy with respect to the set of $\theta_i$. The Gibbs free energy with magnetic field H applied perpendicular to the layer (i.e., along the z-axis) is given as follows:

$$G = - \sum_{i=1}^{N_1+N_2+N_3+N_4-1} \frac{A^{ex}_{i,i+1}}{d_{i,i+1}} cos(\theta_{i+1} - \theta_i) + \sum_{i=1}^{N_1+N_2+N_3+N_4} \left(K_i - \frac{1}{2}\mu_0 M_i^2\right) sin^2(\theta_i) - \sum_{i=1}^{N_1+N_2+N_3+N_4} \mu_0 H M_i cos(\theta_i)$$

in which the first, second and third terms are ferromagnetic exchange, ferromagnetic effective anisotropy and Zeeman energies, respectively and $N_k$ are the number of monolayers in each of the magnetic layers. The indices, k=1,2,…,4 refer to [Co/Pd]$_{4-n}$, [Co/Ni], [Co/Pd]$_n$ and CFB layers, respectively. The layer thickness, $d_i$, exchange stiffness between two nearest-neighbor monolayers, $A_{(i,i+1)}$, magnetocrystalline anisotropy, $K_i$, and saturation magnetization, $M_i$, are used as material specific input parameters; $\theta_i$ is the angle between the z-axis and the magnetization vector within monolayer i. Additionally, we consider the anisotropy constants as effective values that include volume, surface, and interface contributions. The equilibrium state is determined by optimizing the coupled nonlinear equations with the Weierstrass–Erdmann boundary conditions [55].

The [Co/Pd] multilayer is treated as a continuous slab with $A_i^{ex} = 2$ pJ/m, $K_i=0.15$ MJ/m$^3$ and $M_i=0.355$ MA/m for $i \leq N_1$ and $N_2 \leq i \leq N_3$. The last quantity is directly extracted from VSM measurements on a single [Co/Pd]$_5$ multilayer. The value of the exchange stiffness for the Co/Ni layer is estimated to be 12 pJ/m, which is consistent with A ~10 pJ/m for Co-based magnetic thin films. Following reported values [19,42] and the layer thickness dependence of $M_S$ and $K_u$ on 1/t commonly seen in PMA MLs [56,57], we have used saturation magnetization and anisotropy constant equal to 0.75 MA/m and 0.6 MJ/m$^3$, respectively. Material parameters used for the CFB layer are K=0 pJ/m, $M_S=0.625+0.0875 \times t_{CFB}$ (nm) MA/m, and $A_i^{ex} = 19$ pJ/m. Note that we take into account the strong thickness-dependence of $M_S$ for ultra-thin CFB ($t_{CFB} \leq 5.0$ nm). These results are based on ferromagnetic resonance (FMR) measurements of single CFB films, and are in a good agreement with previous results [58]. Accurately modeling this parameter is critical for the simulations because of the strong dependence of the IMA on the CFB thickness.

Fig. 5 summarizes for both sample series A and B the calculated tilt angle, $\theta_M(z)$, of the magnetization through the entire PMA stack thickness and the CFB layer. The results clearly show that the magnetization tilt angle can be engineered by placing the Ni at different locations within the PMA stack (Fig. 5(a)). The tilt angle at the position of the Ni insertion layer becomes larger while being closer to the CFB: angles of 40 degrees, 19 degrees and 8 degrees were calculated for n=0, 1 and 2, respectively. Consistent with the experimental data shown in Fig. 1(b), that is the reduction in PMA as the Ni insertion layer moves deeper into the Co/Pd, we find that the tilt angle of the magnetization of the CFB layer progressively increases with respect to the surface normal: angles of 44°, 63° and 69° were calculated for n=0, 1 and 2, respectively. Moreover, in series B, the simulation results in Fig. 5(b) show that the magnetization tilt angle can be tuned freely as a function of $t_{CFB}$, a trend which was also found in previous works [39,41]. Within the transition region, 0.5 nm $\leq t_{CFB} \leq$ 1.75 nm, a clear tilting of the magnetization $\theta_M$ from 0° to 69° is found, which is consistent with the major loop remanence values. The magnetization configuration is primarily OOP (0°) for $t_{CFB}<1$ nm and gradually turns towards IP with increasing $t_{CFB}$.

In conclusion, we have presented the first experimentally obtained depth profiles of the magnetization in tilted exchange springs using element specific XMCD measurements. We have shown that the magnetization in [Co/Pd]$_{4-n}$-[Co/Ni]-[Co/Pd]$_n$-CFB exchange springs exhibits a strong angle gradient, extending throughout all layers. The achievable tilt angles cover a wide range with the top CFB angle continuously tunable from 0 to 73 degrees by varying its thickness between 0 and 1.75 nm. The magnetization profile can further be tuned by varying the position of the Ni layer in the PMA stack, with a deeper Ni position favoring a steeper gradient and hence a greater CFB angle. Micromagnetic calculations corroborate our experimental results, providing a more detailed description of the magnetization profile, and allow for the tailored design of tilted exchange springs for use STT-MRAM, STOs, and domain wall devices.

**ACKNOWLEDGMENTS**

This work was supported by the EC FP7 Contract ICT-257159 "MACALO", the Swedish Foundation for Strategic Research (SSF), the Swedish Research Council (VR), and the Knut and Alice Wallenberg Foundation. Johan Åkerman is a Royal Swedish Academy of Sciences Research Fellow supported by a grant from the Knut and Alice Wallenberg Foundation. T. N. Anh Nguyen acknowledges financial support from the National Foundation for Science and Technology Development of Vietnam (NAFOSTED) through the Project No. 103.02-2010.27 and from the KIST-IRDA through Alumni Project No. 2Z03750. CWM acknowledges support from the NSF.

**Figure captions**

Fig. 1. (a) Schematic illustration of the control samples to investigate the impact of inserting Ni into the PMA stack. The samples have a constant CFB thickness (0.5 nm) deposited on three different [Co/Pd]$_{4-n}$-Ni-[Co/Pd]$_n$-CFB (n= 0, 1, and 2) stacks and a Ni-free [Co/Pd]$_5$ stack. (b) OOP hysteresis loops showing how the coercivity varies slightly with the presence, and location, of the Ni layer. All measurements were normalized to their saturation magnetization.

Fig. 2. (a) Schematic illustration of the tilted exchange spring material stack with a CFB wedge deposited on three [Co/Pd]$_{4-n}$-[Co/Ni]-[Co/Pd]$_n$ multilayers with different Ni position. The magnetization tilt angle ($\theta_M$) is defined with respect to the film normal. (b) and (c) show IP and OOP hysteresis loops for five different CFB thicknesses. All measurements were normalized to their magnetization value at 1 T.

Fig. 3. Asymmetry of Fe L$_3$ (left column) and Ni L$_3$ absorption edges (right column) for samples in series A. The Fe signal shows a strong IP character which decreases as the Ni layer lies closer to the CFB. The Ni signal changes from OOP to almost IP as it approaches the CFB.

Fig. 4. (a) The magnetization tilt angles ($\theta°$) of Fe, Co, and Ni as the function of $t_{CFB}$ (left) and Ni depth (right), extracted from XMCD spectra. (b) The Co asymmetry for sample CFB(1.75 nm)/n=2 (indicated with a blue solid circles marker in Fig. 4(a)). The solid black line is fitted to the measured asymmetries and contains contributions from $m_S$, $m_D$ and $m_{orb}$. Spin contributions from different Co layers are plotted as green and purple lines, where the spin magnetization direction of the first Co layer was used as a fitting parameter. Sum of all spin contributions is plotted as red solid line.

Fig. 5. The calculated tilt angle, $\theta_M(z)$, of the local magnetization throughout the entire film thickness. (a) Samples from series A, showing how the magnetization profile is strongly affected by the position of the Ni layer: the deeper the Ni position, the steeper the overall gradient and the higher the CFB tilt angle. (b) Samples from series B, showing how the CFB tilt angle can be tuned continuously by varying its thickness.

**Figures**

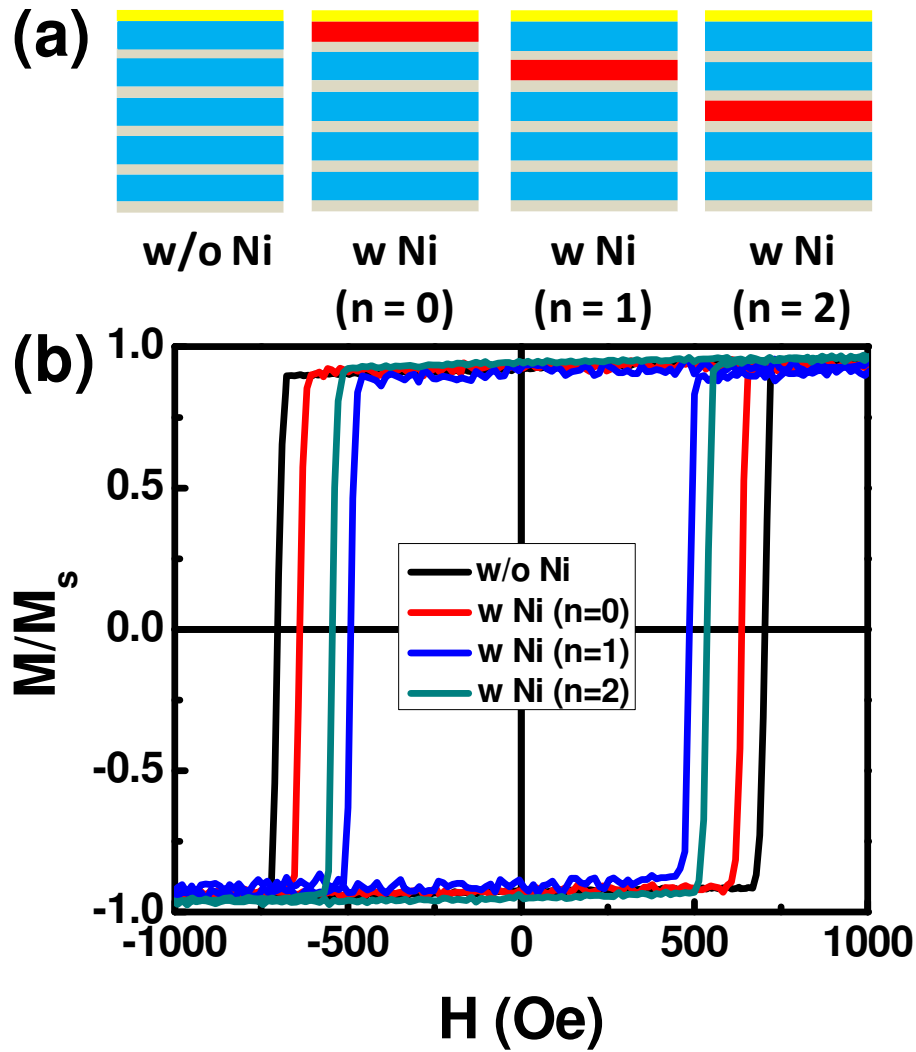

Fig. 1. (a) Schematic illustration of the control samples to investigate the impact of inserting Ni into the PMA stack. The samples have a constant CFB thickness (0.5 nm) deposited on three different $[Co/Pd]_{4-n}$-Ni-$[Co/Pd]_n$-CFB (n= 0, 1, and 2) stacks and a Ni-free $[Co/Pd]_5$ stack. (b) OOP hysteresis loops showing how the coercivity varies slightly with the presence, and location, of the Ni layer. All measurements were normalized to their saturation magnetization.

**Fig.1, Anh Nguyen et al.**

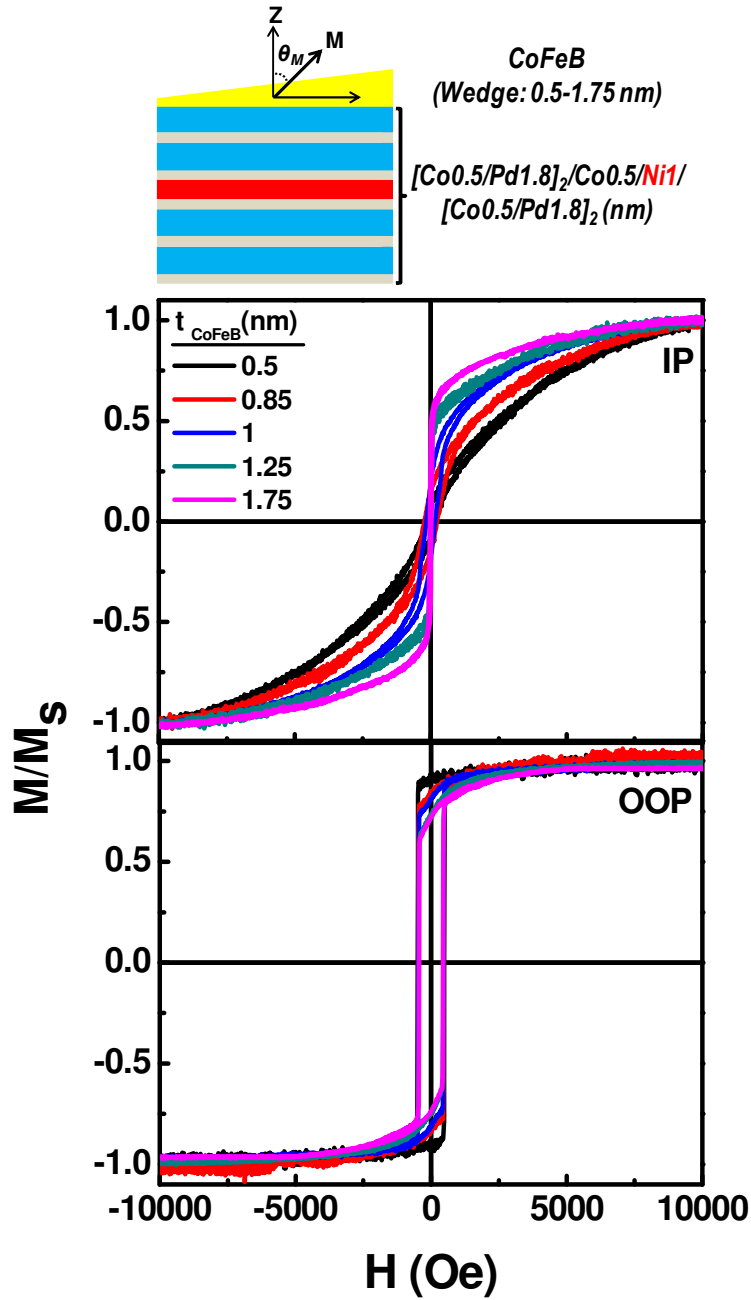

Fig. 2. (a) Schematic illustration of the tilted exchange spring material stack with a CFB wedge deposited on three $[Co/Pd]_{4-n}$-$[Co/Ni]$-$[Co/Pd]_n$ multilayers with different Ni position. The magnetization tilt angle ($\theta_M$) is defined with respect to the film normal. (b) and (c) show IP and OOP hysteresis loops for five different CFB thicknesses. All measurements were normalized to their magnetization value at 1 T.

**Fig.2, Anh Nguyen et al.**

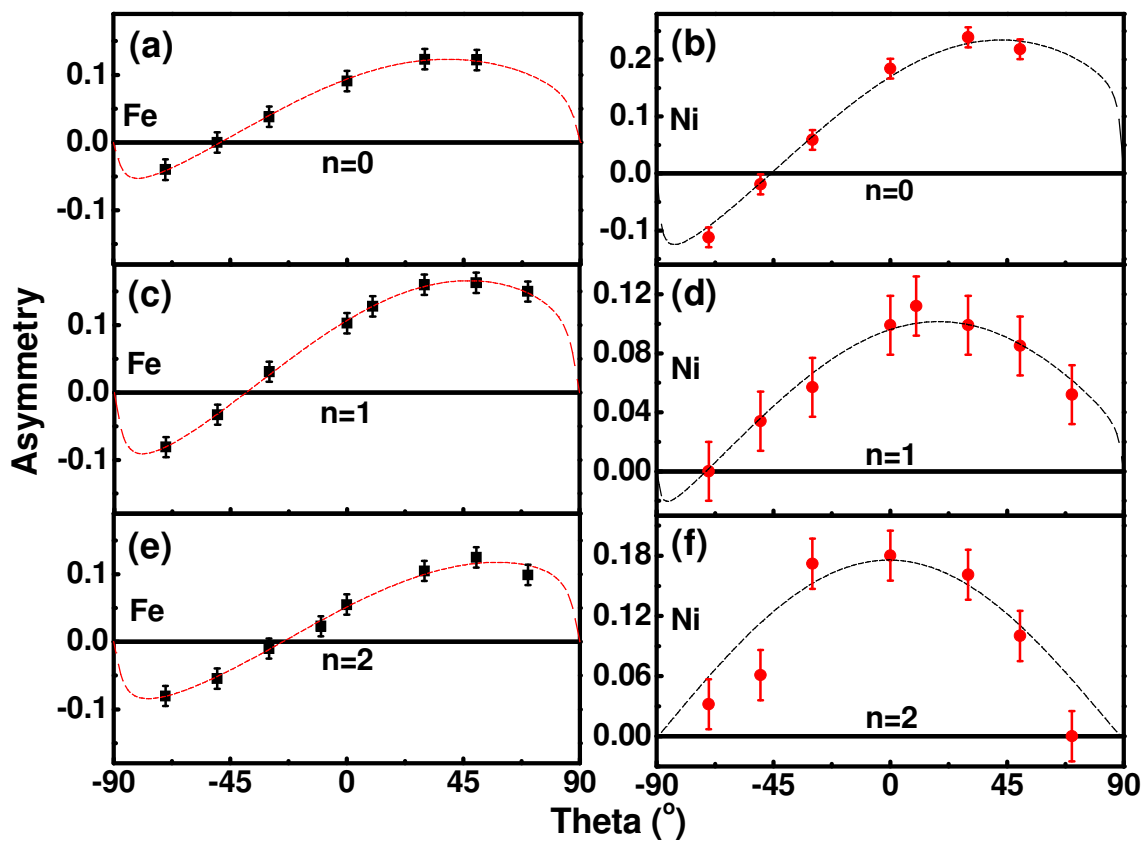

Fig. 3. Asymmetry of Fe L$_3$ (left column) and Ni L$_3$ absorption edges (right column) for samples in series A. The Fe signal shows a strong IP character which decreases as the Ni layer lies closer to the CFB. The Ni signal changes from OOP to almost IP as it approaches the CFB.

**Fig.3, Anh Nguyen et al.**

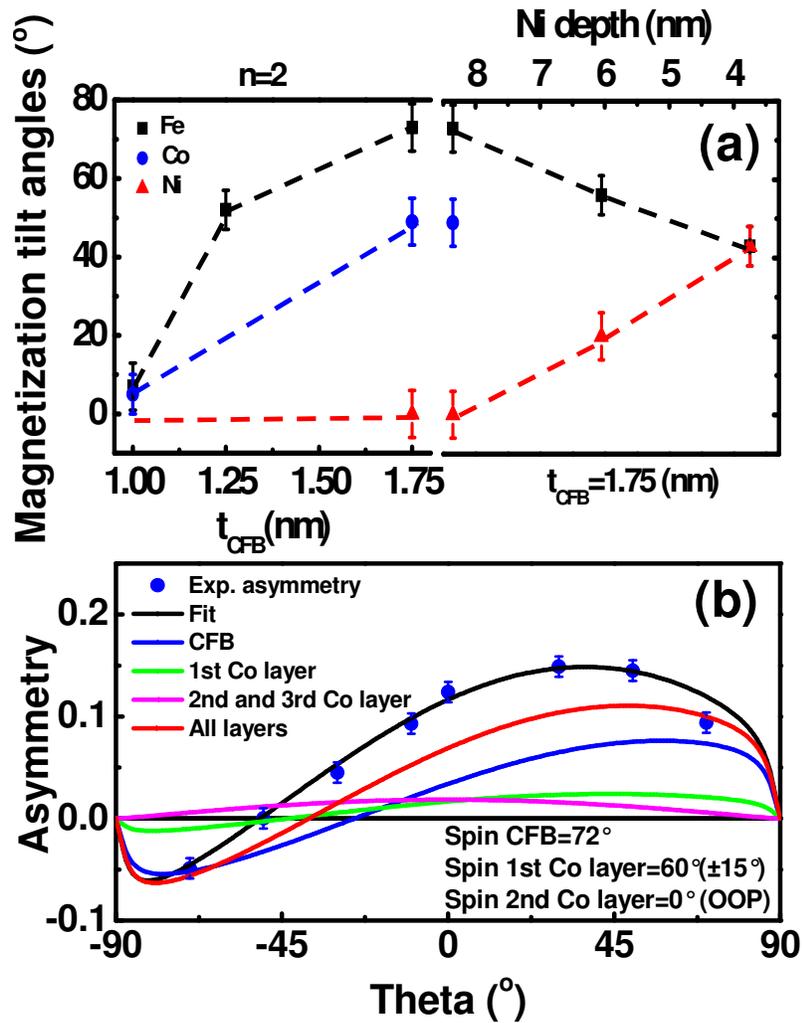

Fig. 4. (a) The magnetization tilt angles (θ°) of Fe, Co, and Ni as the function of $t_{CFB}$ (left) and Ni depth (right), extracted from XMCD spectra. (b) The Co asymmetry for sample CFB(1.75 nm)/n=2 (indicated with a blue solid circles marker in Fig. 4(a)). The solid black line is fitted to the measured asymmetries and contains contributions from $m_S$, $m_D$ and $m_{orb}$. Spin contributions from different Co layers are plotted as green and purple lines, where the spin magnetization direction of the first Co layer was used as a fitting parameter. Sum of all spin contributions is plotted as red solid line.

**Fig.4, Anh Nguyen et al.**

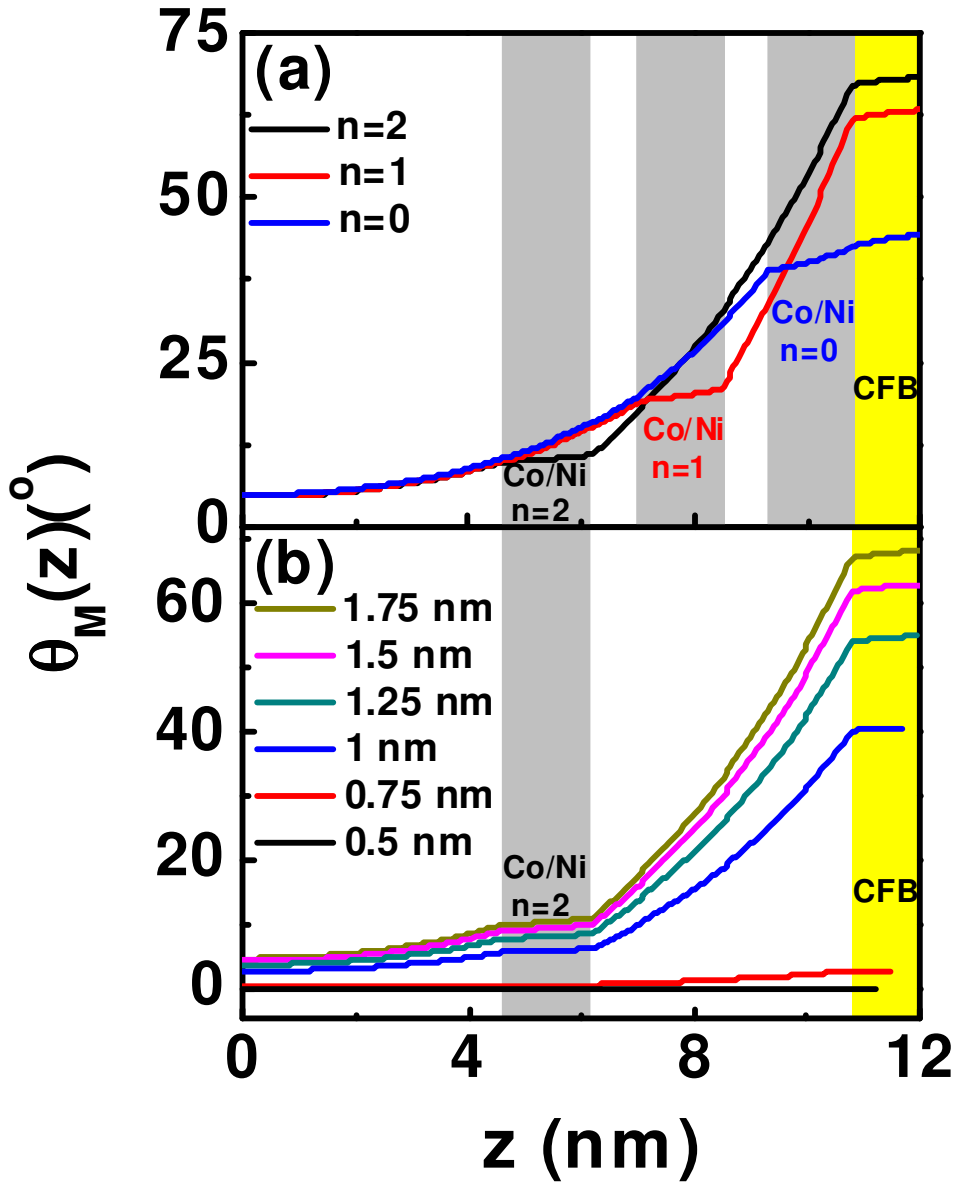

Fig. 5. The calculated tilt angle, $\theta_M(z)$, of the local magnetization throughout the entire film thickness. (a) Samples from series A, showing how the magnetization profile is strongly affected by the position of the Ni layer: the deeper the Ni position, the steeper the overall gradient and the higher the CFB tilt angle. (b) Samples from series B, showing how the CFB tilt angle can be tuned continuously by varying its thickness.

**Fig.5, Anh Nguyen et al.**